\documentstyle[psfig,manuscript,aps]{revtex}
\tightenlines
\begin{document}
\title{Nonquasiparticle states in half-metallic ferromagnet NiMnSb}
\author{L. Chioncel$^{1}$, M. I. Katsnelson$^{1,2,3}$,  
R. A. de Groot$^{1}$, and A. I. Lichtenstein$^{1}$ }
\address{$^1$ University of Nijmegen,NL-6525 ED Nijmegen, The Netherlands\\
$^2$ Institute of Metal Physics, 620219 Ekaterinburg, Russia \\
$^3$ Department of Physics, Uppsala University, Box 530, 
SE-75121 Uppsala, Sweden}

\maketitle

\begin{abstract}
Nonquasiparticle states above the Fermi energy are studied by 
first-principle dynamical mean field calculations for a prototype 
half-metallic ferromagnet, NiMnSb. We present a quantitative evaluation 
of the spectral weight of this characteristic feature and discuss the 
possible experimental investigation (BIS, NMR, STM and Andreev reflection)
to clarify the existence of these states.
\end{abstract}

\section{Introduction}

Half-metallic ferromagnets (HMF) \cite{degroot,IK,pickett} are now a 
subject of growing interest, first of all, because of their possible 
applications to ``spintronics'', that is, spin-dependent electronics 
\cite{prinz}. Being metals for one spin projection and semiconductors 
for the opposite one \cite{degroot} they have order-in-magnitude
different spin contributions to electronic transport properties which 
can result in a huge magnetoresistance for heterostructures containing 
HMF \cite{IK}. In addition to heterostructure systems, bulk materials 
such as La$_{1-x}$Sr$_{x}$MnO$_{3}$  \cite{park} compound, combining 
half-metallic ferromagnetism and colossal magnetoresistance, has also 
attracted more attention to this problem.

As a result, numerous first-principle electronic structure calculations 
of HMF have been carried out, starting from  Ref. \onlinecite{degroot}  
(see, e.g., recent papers \cite{wijs,freeman} and a review of early 
works in Ref. \onlinecite{IK}). All of them are based on a standard 
local density approximation (LDA) or generalized gradient approximation 
(GGA) of the density functional theory or, sometimes, on the LDA+U 
approximation (Ref. \onlinecite{korotin} for CrO$_{2}$). All of these
approaches completely neglect the effects of dynamical spin fluctuations
on the electronic structure which can be of crucial importance for HMF.

The appearance of nonquasiparticle states in the energy gap near the Fermi
level \cite{edwards,IK1} is one of the most interesting correlation effects
typical for HMF. The origin of these states is connected with 
``spin-polaron'' processes: the spin-down low-energy electron excitations, 
which are forbidden for HMF in the one-particle picture, turn out to be 
possible as superpositions of spin-up electron excitations and virtual 
{ magnons}. The density of these nonquasiparticle states vanishes at 
the Fermi level but increases drastically at the energy scale of the order
of a characteristic { magnon} frequency $\omega_{m}$, giving an 
important contribution, in the temperature dependence of the conductivity 
due to the interference with impurity scattering \cite{IK}. It is worthwhile
to mention that the existence of such a nonquasiparticle state is 
important for spin-polarized electron spectroscopy \cite{IK1,KE}, 
NMR \cite{NMR}, and subgap transport in ferromagnet-superconductor 
junctions (Andreev reflection) \cite{falko}.

The temperature
dependence of the HMF electronic structure and stability of half-metallicity
against different spin-excitations are crucial for practical applications
in spintronics. A simple attempt to incorporate the static non-collinear
spin-configurations, due to  finite-temperature effects \cite{spirals},
shows the mixture of spin up and spin down density of states that destroy
the half-metallic behavior. It is our scope to use a more natural 
many-body approach to investigate the proper dynamical spin fluctuations
effect on the electronic structure at temperatures $T<T_c$, within the 
half-metallic ferromagnetic state.

In this paper we present the first quantitative theory of nonquasiparticle 
states in HMF based on realistic electronic structure calculation in 
NiMnSb. The combination of local density approximation in the frame of 
density functional theory with the many-body technique allowed us to 
estimate the spectral weight of the nonquasiparticle states. Various 
possibilities of experimental manifestations of such a states are 
discussed at the end of the paper.

\section{Nonquasiparticle states: an illustrative example}

\begin{figure}
\centerline{\psfig{file=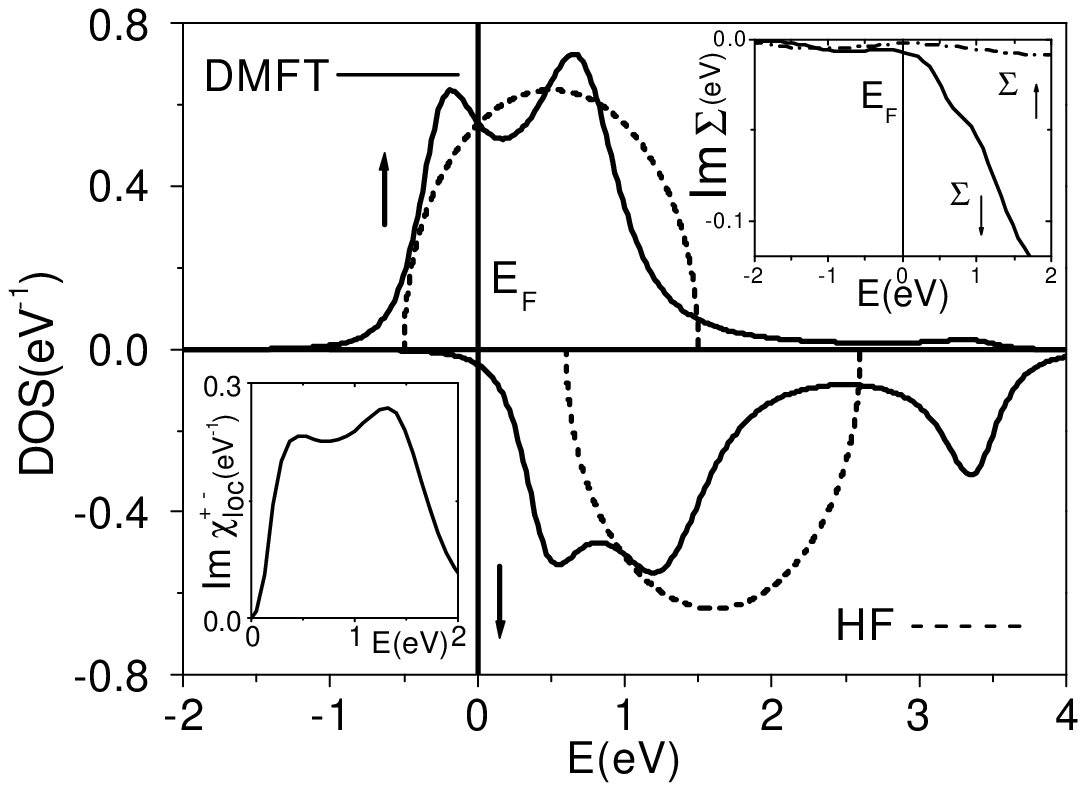,height=3.0in,angle=0}}
\vskip  0.25cm
\caption{
Density of states for HMF in the Hartree-Fock (HF) approximation 
(dashed line) and the QMC solution of DMFT problem for semi-circular model
(solid line) with the band-width $W=2$ eV, Coulomb interaction $U=2$ eV,
spin-splitting $\Delta=0.5$ eV, chemical potential $\mu=-1.5$ eV and temperature 
$T=0.25$ eV. Insets: imaginary part of the local spin-flip susceptibility (left)
and the spin-rezolved selfenergy (right).}
\label{model}
\end{figure}

Before the investigation of the real NiMnSb material it is worthwhile to 
illustrate the correlation effects on the electronic structures of HMF using a 
simple ``toy'' model. The one-band Hubbard model of a saturated ferromagnet
can provide us the simplest model of a half-metallic state:
\begin{equation} \label{ham}
H=-\sum_{i,j,\sigma}t_{ij}(c_{i\sigma}^{\dagger}c_{j\sigma} + 
c_{j\sigma}^{\dagger}c_{i\sigma}) + U \sum_{i}n_{i \uparrow} 
n_{i \downarrow}
\end{equation}
Difficulties in solving the Hubbard model Eq. (\ref{ham}) are well known \cite{edwards}. 
Fortunately there is an exact numerical solution in the limit of infinite dimensionality
or large connectivity called Dynamical Mean Field Theory (DMFT) \cite{GKKR}. 
Following this approach we will consider the Bethe lattice with 
coordination $z \rightarrow \infty$ and nearest neighbor hoping 
$t_{ij}=t/ \sqrt z $. In this case a semicircular density of states is
obtained as function of the effective hopping $t$:
$N(\epsilon)=\frac 1 {2 \pi t^2} \sqrt{ 4t^2 - \epsilon^2 }$.
To stabilize the ferromagnetic solution within the Hubbard model is yet 
another difficult problem. It was proved recently, that the  necessary 
conditions for ferromagnetism is a density of state with large spectral 
weight near the band edges \cite{Ulmke98} and the Hund's rule
coupling for the degenerate case \cite{Vollhardt00}. In our "toy" model in 
order to stabilize the HMF state, we add an external magnetic spin 
splitting term $\Delta=0.5$ eV, which mimic the local Hund polarization 
originated from other orbitals in the real NiMnSb compound. This HMF 
state corresponds to a mean-filed (HF) solution with a LSDA-like DOS,
denoted in Fig. \ref{model} as dashed line. 

DMFT maps the many-body system, Eq.(\ref{ham}), onto the self-consistent
quantum impurity model with the effective action \cite{GKKR}:
\begin{equation}\label{Seff}
S_{eff}=- \int_0^\beta d \tau \int_0^\beta d \tau^{\prime} 
c_{\sigma}^{\dagger}(\tau) {\mathcal G}_{\sigma}^{-1}(\tau - \tau^{\prime})
c_{\sigma}(\tau^{\prime}) + {\int_0^\beta d \tau U n_{\uparrow}(\tau) 
n_{\downarrow}(\tau)}
\end{equation}
The effective medium Green function $ {\mathcal G}_{\sigma}$ 
(Weiss function) is connected with the local Green function  
$ G_{\sigma}$ through the sefconsistency condition:
\begin{equation}\label{slfc}
{\mathcal G}_{\sigma}^{-1}=i \omega+\mu-t^2 G_{\sigma}-1/2 \sigma \Delta
\end{equation} 
where $\omega=(2n+1)\pi T, n=0,\pm1, \pm2, ...$ represents the Matsubara 
frequencies corresponding to a temperature $T$ and $\tau$ is the imaginary time.  
The Green function corresponding to the DMFT effective action, Eq.(\ref{Seff}):
$G_{\sigma}(\tau-\tau^{\prime}) = -<T_{\tau} c_{\sigma}(\tau) c_{\sigma}^{\dagger}(\tau^{\prime})%
>_{S_{eff}}$ have been calculated using the Quantum Monte Carlo 
scheme within the so-called exact enumeration technique,  with the number of time slices 
L=25. $T_{\tau}$ represents the time ordering operator. We would like to 
emphasize that due to the symmetry of the ferromagnetic state the local $G_{\sigma}$ and
the effective medium $ {\mathcal G}_{\sigma}$ Green functions are diagonal
in spin space, even in the presence of the interaction part of the 
effective action Eq.\ref{Seff} which describes the spin-flip scattering process.

The applicability of the local approximation to the problem of existence 
of the nonquasiparticle states has been discussed in 
Ref. \onlinecite{Zarubin}. In this limit it is possible to capture some 
features of the magnetic excitations, namely those which can be described 
by the spin susceptibility $\chi ({\bf q}, i \omega)$ for ${\bf q}= 0$ and
${\bf q}= \pi$ (ferromagnetic and antiferroagnetic long-range order, 
correspondingly)\cite{Jarrell92}. As for the case of a generic ${\bf q}$
it is worthwhile to stress that the accurate description of the {magnon}
spectrum is not important for the existence of the nonquasiparticle states 
and for the proper estimation of their spectral weight, but can be 
important to describe an explicit shape of the density of states "tail" in 
a very close vicinity of the Fermi energy. The DMFT, being an optimal {\it local}
approximation for the electron self-energy \cite{GKKR}, should be
adequate for the description of nonquasiparticle states, because of the
weak momentum dependence of the corresponding contributions to the
electron Green function.

Our model allows to study the magnon spectrum through the two-particle
correlation function which is obtained using the QMC procedure \cite{Jarrell92}.
We calculate the local spin-flip susceptibility:
\begin{equation}\label{chipm}
\chi_{loc}^{+-}(\tau-\tau^{\prime})= <S^+ (\tau) S^-(\tau^{\prime})> = 
<T_{\tau} c_{\uparrow}^\dagger(\tau) c_{\downarrow}(\tau) 
c_{\downarrow}^{\dagger}(\tau^{\prime})c_{\uparrow}(\tau^{\prime})>_{S_{eff}}
\end{equation} 
which gives us information about the integrated magnon spectrum \cite{IK1,Moriya85}.

The DMFT results are presented in Fig. \ref{model}. In comparison with a 
simple Hartree-Fock solution one can see an additional well-pronounced 
feature appearing in the spin-down gap region, just above the Fermi level. 
This new many-body feature corresponds to the so called nonquasiparticle 
states in HMF \cite{edwards,IK1} and represents the spin-polaron process 
\cite{edwards,IK1}: the spin-down electron excitations forbidden in the one
electron description of HMF are possible due to the superposition of 
spin-up electron excitations and virtual {magnons}. In addition to this
nonquasiparticle states visible in both spin channels of DOS around 0.5 eV,
a many-body satellite appears at 3.5 eV. 

The left inset of Fig. \ref{model}, represents the imaginary part of local
spin-flip susceptibility. One can see a well pronounced shoulder ($\simeq 0.5$ eV), 
which is related to a characteristic  magnon excitation \cite{IK1}. In 
addition there is a broad maximum ($\simeq 1$ eV) corresponding to the Stoner 
excitation energy.  
The right inset of Fig. \ref{model}, represents the imaginary part of 
self-energy calculated from our "toy model". The spin up channel can be
described by a Fermi-liquid type behavior, with a parabolic energy
dependence $-Im \Sigma^{\uparrow} \simeq (E-E_F)^2$, where as in the 
spin down channel, of $\Sigma^{\downarrow}$, the non-quasiparticle shoulder at 0.5 eV, 
is visible. Due to the relatively high
temperature ($T=0.25$eV) in our QMC calculation the nonquasiparticle tail goes below the
Fermi level. At zero temperature ($T=$0) the tail should end exactly at the
Fermi level \cite{IK1}. Since the exact enumeration technique not "suffer" 
form the QMC noise, the Pade analytical continuation was 
used to extract spectral functions and density of states \cite{pade}.

The existence of the nonquasiparticle states for this model has been
proven by perturbation-theory arguments \cite{edwards} (i.e. for a 
broad-band case) and in the opposite infinite-$U$ limit \cite{IK1}.
Physically, the appearance of these states can
be considered as a kind of spin-polaron effect. According with the 
conservation laws, in the many-body theory the spin-down state with the 
quasimomentum ${\bf k}$ can form a superposition with the spin-up states 
with the qausimomentum ${\bf k-q}$ plus a {magnon} with the 
quasimomentum ${\bf q}$, ${\bf q}$ running the whole Brillouin zone. 
Taking into account the restrictions from the Pauli principle (an 
impossibility to scatter into occupied states) one can prove that this 
superposition can form only above the Fermi energy (here we consider the
case where the spin-up electronic structure is metallic and the spin-down
is semiconducting; oppositely, the nonquasiparticle states form only 
{\it below} the Fermi energy) \cite{IK,IK1}. If we neglect the {magnon}
energy in comparison with the typical electron one than the density of 
nonquasiparticle states will vanish abruptly right at the Fermi energy; 
more accurate treatment shows that it vanishes continuously in the interval
of the order of the {magnon} energy with a law which is dependent on 
the {magnon} dispersion \cite{edwards}. As a consequence          
the nonquasiparticle states are almost currentless \cite{IK,IK1}. Recently,
some evidences of the existence of almost currentless states near the Fermi
energy in half-metallic ferromagnet CrO$_{2}$ have been obtained by x-ray
spectroscopy \cite{kurmaev}. 

\section{Computational method}

Recently, an approach to include correlation effects into the first-principle
electronic structure calculations by the combination of the LDA with dynamical
mean-field theory, DMFT (for review of DMFT, see Ref.\onlinecite{GKKR})
has been proposed \cite{AnisDMFT,lda++}. In this case the DMFT maps a 
lattice many-body system onto multi-orbital impurity model subject to a
self-consistent condition in such a way that the many-body problem
splits into one-body impurity problem for the crystal and the many-body 
problem for an effective atom. Therefore, the approach is complementary
to the local (spin) density approximation \cite{hohenberg64,kohn65,hedin} 
where the many-body problem splits into one-body problem for a crystal and
many-body problem for {\it homogeneous} electron gas. Naively speaking, the
LDA+DMFT method \cite{AnisDMFT,lda++} treats $d$- and $f$-electrons in 
spirit of DMFT and $s,p$-electrons in spirit of LDA. Of course, this is a 
crude description since these two subsystems are not considered as 
independent ones but connected by the self-consistency conditions. In fact,
the DMFT, due to numerical and analytical techniques developed to solve the
effective impurity problem \cite{GKKR}, is a very efficient and extensively
used approximation for energy dependent self-energy $\Sigma (\omega)$. The
emerged LDA+DMFT method can be used for calculating a large number of
systems with different strength of the electronic correlations (for 
detailed description of the method and computational results, see 
Refs. \onlinecite{lichtenstein01,held02,kotliar01}). The LDA+DMFT
method appeared to be efficient in the consideration of a series of 
classical problems which were beyond the standard density functional theory,
for example, electronic structure of the Mott-Hubbard insulators 
\cite{voll}, magnetism of transition metals at finite temperatures 
\cite{FeNi} and $\alpha$-$\delta$ transition in Pu \cite{Pu}. Here we 
present the results of LDA+DMFT calculations of the electronic 
structure of a ``prototype'' half-metallic ferromagnet NiMnSb.

In order to integrate the dynamical mean field approach into the band 
structure calculation we use the so called exact muffin-tin orbital 
method (EMTO)\cite{Andersen,EMTO}. In our current implementation 
\cite{EMTODMFT}, in addition to the usual self-consistency of the 
many-body problem (self-consistency of the self-energy), we also 
achieved charge self-consistency. In the EMTO approach the one electron 
effective potential is represented by the optimized overlapping muffin-tin
potential\cite{Andersen,EMTO}, which is the best possible spherical 
approximation to the full one-electron potential. The effective potential
is used to calculate the one-electron Green function $G_{EMTO}({\bf  k},z)$,
on an arbitrary complex energy contour $z$, which encloses the valence band
poles of the one-electron Green function. For core electrons a frozen core
approach is used. For any Bloch wave vector ${\bf k}$ from the Brillouin
zone and complex energy $z$, the local multi-orbital self-energy $\Sigma(z)$
is added to the LDA Green function via the Dyson equation \cite{EMTODMFT}:
\begin{equation}\label{GDMFT}
G(z)=\sum_{\bf  k}\left[ G_{EMTO}^{-1}({\bf  k},z) -\Sigma(z) \right]^{-1}
\end{equation}
(all the quantities here are matrices in spin, orbital, and, for several 
atoms per unit cell, site indices). In the iteration procedure the LDA+DMFT
Green function (\ref{GDMFT}) is used to calculate the charge and spin 
densities. Finally, for the charge self-consistency calculation we 
construct the new LDA effective potential from the spin and charge 
densities \cite{EMTODMFT}, using the Poisson equation in the spherical cell
approximation \cite{vitos01}.

For the interaction Hamiltonian, we have taken the most general 
rotationally invariant form of the generalized Hubbard (on-site) 
Hamiltonian \cite{lda++}. The many-body problem is solved using 
the SPTF method proposed in Ref. \cite{Katsnelson02}, which is a 
development of the earlier approach \cite{lda++}. The SPTF approximation is
a multiband spin-polarized generalization of the fluctuation exchange
approximation (FLEX) of Bickers and Scalapino, but with a different
treatment of particle-hole (PH) and particle-particle (PP) channels. 
Particle-particle (PP) channel is described by a $T$-matrix approach 
\cite{galitski63} giving a renormalization of the effective interaction. 
This effective interaction is used explicitly in the particle-hole channel.
Justifications, further developments and details of this scheme can be 
found in Ref.\ \cite{Katsnelson02}. Here we present the final expressions
for the electron self-energy. The sum over the ladder graphs leads to the
replacement of the bare electron-electron interaction by the $T$-matrix 
which obeys the equation:

\begin{equation}\label{tmat}
<13|T^{\sigma \sigma ^{\prime }}(i\Omega)|24>=<13|v|24>-T\sum_{\omega}
\sum_{5678}{\cal G}_{56}^{\sigma }(i\omega )
{\cal G}_{78}^{\sigma ^{\prime }}(i\Omega -i\omega )
<68|T^{\sigma \sigma ^{\prime }}(i\Omega)|24>
\end{equation}
where the matrix elements of the screened Coulomb interaction,
$<13|v|24>$, are expressed using the average Coulomb and exchange
energies $U,J$ \cite{Katsnelson02}. $|1>=|j,m>$ where $(j)$ is the site-number, 
$(m)$ the orbital quantum number,  $\sigma, \sigma \prime$ are the
spin indices, $T^{\sigma \sigma ^{\prime }}(i\Omega)$ represents the $T$-matrix 
and T is the temperature. In the following we write the
perturbation expansion for the interaction (\ref{tmat}). The two 
contributions to the self-energy are obtained by replacing of the bare
interaction by a $T$-matrix in the Hartree and Fock terms:

\begin{eqnarray}\label{thtf}
\Sigma_{12}^{\sigma,{\rm TH}}(i\omega )&=&T\sum_{\Omega}
\sum_{34\sigma^{\prime}} <13|T^{\sigma\sigma^{\prime}}(i\Omega)|24>
{\cal G}_{43}^{\sigma^{\prime}}(i\Omega-i\omega) \nonumber\\
\Sigma _{12}^{\sigma,{\rm TF}}(i\omega )&=&-T\sum_{\Omega}
\sum_{34\sigma^{\prime}}<14|T^{\sigma\sigma^{\prime}}(i\Omega)|32>
{\cal G}_{34}^{i\sigma^{\prime}}(i\Omega-i\omega)
\end{eqnarray}
The four matrix elements of the bare longitudinal susceptibility
represents the density-density $(dd)$, density-magnetic $(dm^{0})$,
magnetic-density $(m^{0}d)$ and magnetic-magnetic channels $(m^{0}m^{0})$.
The matrix elements couples longitudinal magnetic fluctuation with density
magnetic fluctuation. In this case the particle-hole contribution to the
self-energy is written in the Fourier transformed form

\begin{equation}\label{selfph}
\Sigma _{12}^{\sigma,{\rm PH}}(\tau )=\sum_{34\sigma ^{\prime }}
W_{1342}^{\sigma \sigma ^{\prime }}(\tau )
{\cal G}_{34}^{\sigma ^{\prime }}(\tau )
\end{equation}
The particle-hole fluctuation potential
matrix $W^{\sigma \sigma ^{\prime }}(i\omega )$ is defined in the FLEX
approximation \cite{bicker89,lda++} with the replacement of the bare 
interaction by the ``static'' $T$-matrix. The 
effective particle-hole fluctuation potential is an energy dependent
quantity and is determined self-consistently:

\begin{equation}\label{W}
W^{\sigma \sigma ^{\prime }}(i\omega )=\left(
\begin{array}{cc}
{W}_{\uparrow \uparrow}(i\omega )&{W}_{\uparrow \downarrow }(i\omega ) \\
{W}_{\downarrow \uparrow}(i\omega ) & W_{\downarrow \downarrow }(i\omega )
\end{array}
\right)
\end{equation}

As can be seen from expression Eq.(\ref{W}), the spin polarized T-matrix FLEX
approach describe the spin-flip scatterings corresponding to the non-diagonal part 
of the $W^{\sigma \sigma ^{\prime}}(i\omega)$ matrix \cite{Katsnelson02}. Nevertheless,
the local and Weiss Green functions as well as the electronic
self-energies are spin diagonal, due to the symmetry of 
the ferromagnetic state. 

It is important to note that all of the above expressions for the self-energy, in
the spirit of the DMFT approach, involve the Weiss Green function. 
The total self-energy is obtained from Eqs.\ (\ref{thtf}) and (\ref{selfph}):

\begin{equation}
\Sigma^{\sigma}(i\omega)=\Sigma^{\sigma,{\rm TH}}(i\omega)+
\Sigma^{\sigma,{\rm TF}}(i\omega)+\Sigma^{\sigma,{\rm PH}}(i\omega).
\end{equation}

Since some part of the correlation effects are included already in
the local spin-density approximation (LSDA) a ``double counted'' terms
should be taken into account. To this aim, we start with the LSDA
electronic structure and replace 
$\Sigma_{\sigma}(E)$ by $\Sigma_{\sigma}(E)-\Sigma_{\sigma}(0)$ in all 
equations of the LDA+DMFT method. It means that we only add 
{\it dynamical} correlation effects to the LSDA method.

We would like to emphasize that Eq.(\ref{selfph})
includes spin flip scattering missing from the standard GW approach,
and these processes are responsible for the appearance of
spin-polaron, or nonquasiparticle, states in the energy gap of HMF.
The $T$-matrix renormalization is important for proper description
of these processes which can be demonstrated accurately for the Hubbard
\cite{IK1} and s-d exchange \cite{irkhin85} models in the spin-wave 
temperature region; in both cases it is the $T$-matrix (and not the 
bare interaction) that determines the amplitudes of {electron-magnon}
interactions.

\section{Results: N\lowercase{i}M\lowercase{n}S\lowercase{b}}

In our LDA calculations we considered the standard representation of the
$C1_b$ structure with a fcc unit cell containing three atoms: $Ni(0,0,0)$, 
$Mn(1/4,1/4,1/4)$, $Sb(3/4,3/4,3/4)$ and a vacant site $E(1/2,1/2,1/2)$
respectively. We used the experimental lattice constant of NiMnSb 
($a=5.927 \AA$) for all the calculations. To calculate the charge
density we integrate along a contour on the complex energy plane which
extends from the bottom of the band up to the Fermi level \cite{EMTO}, 
using 30 energy points. For Brillouin zone integration we sum up a k-space
greed of 512 points in the irreducible part of the Brillouin zone. A cutoff
of $l_{max}=8$ for the multipole expansion of the charge density and a 
cutoff of $l_{max}=3$ for the wave functions was used. The Perdew-Wang \cite{PW}
parameterization of the Local Density Approximation to the exchange 
correlation potential was used.

\begin{figure}
\centerline{\psfig{file=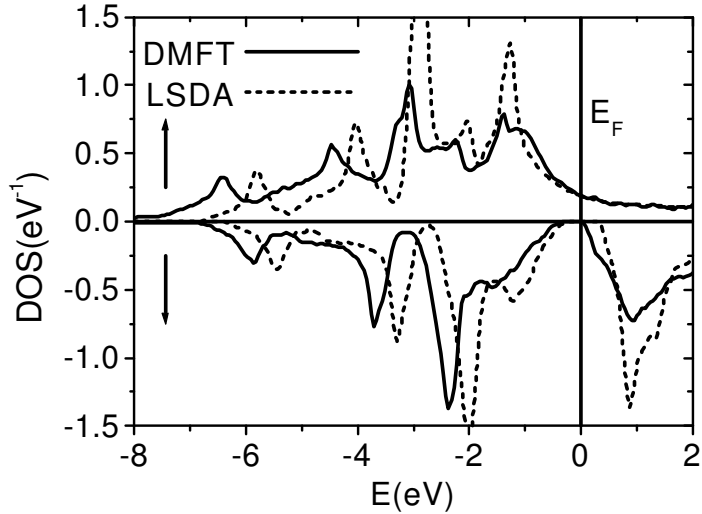,height=3.0in}}
\vskip  0.25cm
\caption{Density of states for HMF
NiMnSb in LSDA scheme (dashed line) and in LDA+DMFT scheme (solid line)
with  effective Coulomb interaction $U$=3 eV, exchange parameter $J$=0.9 eV
and temperature $T$=300 K. The nonquasiparticle state is evidenced just
above the Fermi level.}
\label{nimnsb}
\end{figure}

In order to incorporate the DMFT approach into the realistic electronic
structure we need to evaluate the average on-site Coulomb repulsion energy $U$ 
and the exchange interaction energy $J$. We used the constrained LDA calculation
\cite{AnisimovU} which gives the value of the average Coulomb interaction between the Mn $d$ 
electrons equal to $U=4.8$ eV and the exchange interaction energy equal to $J=0.9$ eV. 
Because the $3d$ orbitals of Ni are fully occupied, correlation effects 
are not so important. The insulating screening used in the constraint LDA-calculation
\cite{AnisimovU} should be generalized to a metallic one as in the case of 
HMF. Such a generalization will lead to additional reduction of the  value of $U$.
Therefore, we performed LDA+DMFT calculations for the different values of $U$ between
$0.5$ and $4.8$ eV. On the other hand, the results of constrained LDA calculations for the Hund
exchange parameter $J$ are not sensitive to the metallic screening \cite{Anisimovrev}.
Our LDA+DMFT results shows a very weak $U$ dependence, due to the $T$-matrix
renormalization \cite{Katsnelson02}. Fig. \ref{nimnsb} represents the typical results 
for density of states using the values of $U=3$ eV, and $J=0.9$ eV.
In comparison with the LDA the LDA+DMFT density of states shows the existence of 
new states in the LDA gap of the spin down channel just above the Fermi level. 

It is important to mention that the magnetic moment per formula unit is not
sensitive to the $U$ values. For a temperature equal to $T=300K$ the calculated 
magnetic moment, $\mu=3.96 \mu_B$, is close to the integer LDA-value $\mu=4.00 \mu_B$, 
which suggests that the half-metallic state is stable with respect to the introduction 
of the correlation effects. In addition, the DMFT gap in the spin down channel, 
defined as the distance between the occupied part and starting point of 
nonquasiparticle state's "tail", is also not very  sensitive to the $U$ 
values. For different $U$ the slope of the ``tail'' is slightly changed,
but the total DOS is weakly $U$-dependent due to the $T$-matrix 
renormalization effects.

\begin{figure}
\centerline{\psfig{file=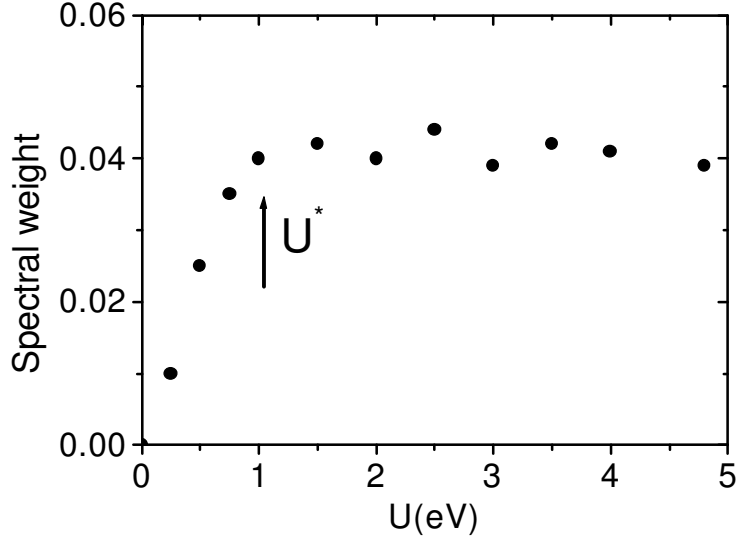,height=3.0in}}
\vskip  0.25cm
\caption{Spectral weight of the nonquasiparticle state, calculated as
function of average on-site Coulomb repulsion $U$ at temperature $T$=300 K.}
\label{spec}
\end{figure}

Thus the correlation effects do not effect too strongly on a general
picture of the electron energy spectrum (except the smearing of the 
density of states features which is due to the finite temperature $T=300K$
in our calculations). The only qualitatively new effect is the appearance
of the nonquasiparticle states in the energy gap above the Fermi
energy. Their spectral weight for realistic values of the parameters are
not too small which means that they should be well-pronounced in the
corresponding experimental data. A relatively weak dependence of the 
nonquasiparticle spectral weight on the $U$ value (Fig. \ref{spec}) is also
a consequence of the $T$-matrix renormalization \cite{Katsnelson02}.

One can see that the $T$-matrix is slightly dependent on $U$ provided that
the latter is larger than the widths of the main density of states peaks
situated near the Fermi level (which is of the order of $U^{*}\simeq 1$ eV)
in an energy range of $2$ eV.

\begin{figure}
\centerline{\psfig{file=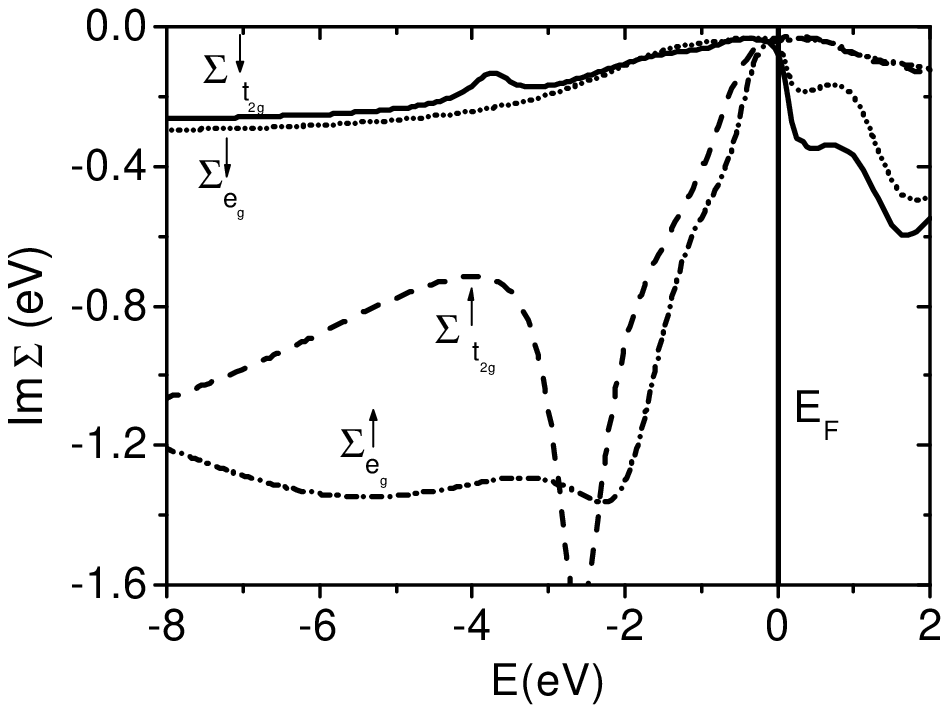,height=3.0in}}
\vskip  0.25cm
\caption{The imaginary part of self-energies $ Im\Sigma_d^{\downarrow}$ for
$ t_{2g}$ (solid line) and $e_g$ (dotted line), $ Im\Sigma_d^{\uparrow}$ 
for $t_{2g}$ (dashed line) and $e_g$ (dashed dotted line) respectively.}
\label{self}
\end{figure}

For spin-up states we have a normal Fermi-liquid behavior
$-Im\Sigma_d^{\uparrow}(E) \propto (E-E_F)^2$, with a typical energy
scale of the order of several eV. The spin-down self energy behaves in
a similar way below the Fermi energy, with a bit smaller energy scale
(which is still larger than 1 eV). At the same time, a significant increase
in $Im\Sigma_d^{\downarrow}(E)$ with much smaller energy scale (tenths of
eV) is evidenced right above the Fermi level which is more pronounced for
$t_{2g}$ states (Fig. \ref{self}). The nonquasiparticle states are visible in the
spin $\downarrow$ DOS Fig. \ref{nimnsb}, as well as in the spin $\downarrow$
channel of the imaginary part of $\Sigma^{\downarrow}$, at the same energy. The similar 
behavior is evidenced in the model calculation Fig. \ref{model}.

According to the model consideration \cite{IK,edwards,IK1} the width of
this ``jump'' should be of the order of characteristic {magnon} 
energy which is much smaller than a typical electron band energy scale.
In the simplest case of neglecting the dispersion of the {magnon}
frequency, $\omega _{\mathbf{q}}\approx \omega _{\mathbf{m}}$ 
with respect to the electron hopping energy $t_{\mathbf{k}}$ 
the electronic self-energy becomes local \cite{edwards}:
\begin{eqnarray*}
\Sigma _{\mathbf{k},\downarrow }(E) &=&\frac{U^{2}m}{N}
\sum_{\mathbf{k}^{\prime }}\frac{1-f(\mathbf{k}^{\prime }\uparrow )}
{E-t_{\mathbf{k}^{\prime}\uparrow}+\omega_{\mathbf{k}-\mathbf{k}^{\prime}}
+i\delta }\simeq \\
&\simeq &\frac{U^{2}m}{N}\sum_{\mathbf{k}^{\prime }}
\frac{1-f(\mathbf{k}^{\prime }\uparrow )}
{E-t_{\mathbf{k}^{\prime }\uparrow }+\omega _{\mathbf{m}}
+i\delta }=\Sigma _{\downarrow }^{loc}(E)
\end{eqnarray*}
where $f(\mathbf{k}^{\prime }\sigma )$ is the Fermi distribution function.   
Therefore our main results: (\textit{i}) the existence of the 
nonquasiparticle states in real electronic structure of a specific 
compound, and (\textit{ii}) estimation of their spectral weight, can be 
obtained in the local LDA+DMFT approximation. The nonquasiparticle peak
in the density of states  (Fig. \ref{nimnsb}) is proportional to the 
imaginary part of the self-energy (Fig. \ref{self}), therefore it is 
determined by the processes of quasiparticle decay, which justifies the 
term ``nonquasiparticle'' itself.

\section{Discussion and conclusions}

From the point of view of the many-body theory, the general approach in 
the DMFT is to neglect the momentum-dependence in the electron self-energy.
In many cases such as the Kondo effect, the Mott metal-insulator transition, 
etc. the energy dependence of the self-energy is obviously much more 
important than the momentum dependence and, therefore, the DMFT is
adequate to consider these problems \cite{GKKR}. As for itinerant electron
ferromagnetism, the situation is not completely clear. Note, however, that
the LDA+DMFT treatment of finite temperature magnetism and electronic 
structure in Fe and Ni appeared to be quite successful \cite{FeNi}.
Experimentally, even in itinerant electron paramagnets close to 
ferromagnetic instability, such as Pd, the momentum dependence of the
self-energy does not look to be essential \cite{Pd}. One can expect that
in magnets with well defined local magnetic moments such as half-metallic
ferromagnets local approximation for the self-energy (i.e., the DMFT)
should be even more accurate. In particular, as we discussed above, it
can be used for the calculations of spin-polaronic (nonquasiparticle) 
effects in these materials. 

Several experiments could be performed in order to clarify the impact of
these nonquasiparticle states on spintronics. Direct ways of observing the
nonquasiparticle states would imply the technique of Bremsstrahlung 
Isohromat Spectroscopy (BIS) \cite{BIS} or spin-polarized scanning 
tunneling microscopy \cite{STM}. In contrast with the photoelectron 
spectroscopy (spectroscopy of the occupied states) which show a complete
spin polarization in HMF \cite{park}, BIS spectra should demonstrate an
essential depolarization of the states above $E_{F}$, on the other hand
SP-STM should also be able to probe these states which give the 
minority-spin contribution to the differential tunneling conductivity 
$dI/dV$\cite{Mahan,tun}. Another way to observe the nonquasiparticle 
states is the low-temperature measurement of the longitudinal nuclear 
magnetic relaxation rate $1/T_{1}$. Since the Korringa contribution due to
the Fermi contact hyperfine interaction, $1/T_{1}\propto TN_{\downarrow}
(E_{F})N_{\uparrow}(E_{F})$ vanishes for HMF a specific dependence, 
$1/T_{1}\propto T^{5/2}$ \cite{IK1} should take place \cite{NMR1}. 
Andreev reflection spectroscopy using the tunneling junction 
superconductor - HFM \cite{falko} can also be used in searching the
experimental evidence of the nonquasiparticle states. Finally, we mention
the spin-polarized STM techniques as a possible method of direct
observation of the nonquasiparticle state in half-metallic ferromagnets. 
The spin-polarized scanning tunneling spectroscopy with positive bias
voltage can in principle detect the opposite-spin state just above the
Fermi level for surface of HMF such as CrO$_2$. This experimental
measurements will be of crucial importance for the theory of spintronics in
any tunneling devices with half-metallic ferromagnets. In particular, $I-V$
characteristics of half-metallic tunnel junctions for the case of 
antiparallel spins are completely determined by the nonquasiparticle
states \cite{IKlast}. Keeping in mind that ferromagnetic semiconductors
can be considered as a peculiar case of HFM \cite{IK}, an account of these
states can be important for proper description of spin diodes and
transistors \cite{vignale,falko}. Thus, the realistic computation of the
spectral weight of nonquasiparticle states can be an interesting and
important application of the LDA+DMFT approach.
 
\section{Acknowledgment}           

This work is supported by the Stichting for Fundamenteel 
Onderzoek der Materie (FOM-NWO), Nederladse Organizatie
voer Wetenschappelijk Onderzoek (NWO) project 047-008-16.
The authors thank to G.A. de Weijs, L.Vitos, I. A. Abrikosov  
and O. Eriksson for fruitful discussions.

\begin{references}
\bibitem{NMR1} Note that weaker dipole-dipole interactions result in a
``residual'' linear-in-$T$ relaxation term but, nevertheless, the
nonquasiparticle contribution $T^{5/2}$ should be easily observable; for
details, see Ref. \cite{NMR}.
\end{references}
\end{document}